\documentclass[aps,twocolumn,showpacs,floatfix]{revtex4}

\usepackage{dcolumn}
\usepackage{graphicx}
\usepackage{amsfonts}
\usepackage{amsmath,bm,amssymb}
\usepackage{comment}
\usepackage[usenames]{color}

\begin{document}

\title{Structure, strain, and control of ground state property in
  LaTiO$_3$/LaAlO$_3$ superlattice}

\author{Alex Taekyung Lee} \affiliation{Department of Physics, Korea Advanced 
Institute of Science and Technology (KAIST), Daejeon 305-701, Korea
}

\author{Myung Joon Han} \email{mj.han@kaist.ac.kr}
\affiliation{Department of Physics and KAIST Institute for the
  NanoCentury, Korea Advanced Institute of Science and Technology,
  Daejeon 305-701, Korea }

\date{\today }

\begin{abstract}
Using first-principles density functional theory calculations, we
examined the ground state property of LaTiO$_3$/LaAlO$_3$
superlattice. Total energy calculations, taking account of
the structural distortions, $U$ dependence, and exchange correlation
functional dependence, show that the spin and orbital ground state can
be controlled systematically by the epitaxial strain. In the wide
range of strains, ferromagnetic spin and antiferro orbital ordering are
stabilized, which is notably different from the previously reported
ground state in titanate systems. By applying large tensile strains,
the system can be transformed into an antiferromagnetic spin and
ferro-orbital-ordered phase.
\end{abstract}


\maketitle

\section{Introduction}
Transition-metal oxide (TMO) heterostructures have attracted great
attention in recent years due to their intriguing material
characteristics and potential applications \cite{MRS,Hwang-review}.
Unexpected interface phenomena caused by the heterostructuring and the
band structure change have been reported such as magnetism
\cite{Ohtomo-LAO/STO,Nakagawa,Brinkman,JSLee,sup-LAOSTO-1,sup-LAOSTO-2}
and superconductivity \cite{sup-LAOSTO-1,sup-LAOSTO-2,Reyren}.
Importantly, the material property can be altered by the epitaxial
strain through the strong couplings in TMO between the charge, spin,
orbital, and lattice degrees of freedom \cite{MIT-RMP,Dagotto}. For
example, previous studies have shown that the superconducting
\cite{Locquet,Bozovic}, ferromagnetic (FM) \cite{JZhang}, and
metal-insulator transition temperature \cite{Cao,Razavi,Ahn} can be
controlled by the strain.

LaTiO$_3$/LaAlO$_3$ (LTO/LAO) is an example that shows a
significantly different electronic structure due to the quantum
confinement \cite{Seo}. A recent in-plane and out-of-plane optical
conductivity measurement in combination with LDA+$U$ calculation
demonstrated that the electronic strucutre of a classical Mott
insulator, LTO, is significantly changed by making a heterostructure. In
this study by Seo et al., antiferromagnetic (AFM) spin and ferro
orbital order (OO) with one electron occupying Ti-$d_{xy}$ has been
suggested as the ground state configuration \cite{Seo}. The lifted
degeneracy and the low-lying $d_{xy}$ band caused by the translational
symmetry breaking along the $c$-axis also play an important role in
SrTiO$_3$/LaAlO$_3$ (STO/LAO), which has been actively studied recently
\cite{sup-LAOSTO-1,sup-LAOSTO-2,JSLee}.

We note that the effect of strain and structural distortion have not
been examined in detail in the previous studies of titanate
superlattices. Most calculations, for example, do not consider the
rotational degrees of freedom of the oxygen octahedra around the Ti
ions in the superlattices \cite{Seo,Popovic,
  Okamoto-1,Pentcheva-LTO/STO}.  Importantly, however, the structure
and strain can play crucial roles in determining the ground state spin
and orbital configuration. Considering the structural property of bulk
LTO, the possible rotation and distortion should be investigated
carefully.

In this work, we examined LTO/LAO superlattice using first-principles
density functional methods. Our calculations show that the spin and
orbital ground state, which are qualitatively different from the
previously reported $d_{xy}$-AFM phase, can be stabilized. Due to the
interplay between spin, orbital, and lattice degrees of freedom, the
FM spin and antiferro orbital order is stabilized in a wide range of
strains while the AFM and ferro OO can be realized by applying
$\sim2.8\%$ of tensile strains. Our results suggest a possible way of
controlling the ground state properties of TMO superlattices. \\[1mm]

\section{Calculation method}
The projector augmented wave (PAW) potentials \cite{PAW} are adopted
in our calculations. For the exchange-correlation functional, we used
both the generalized gradient approximation (GGA) proposed by Perdew
et al. (PBE) \cite{PBE} and the revised PBE (PBEsol) \cite{PBEsol} as
implemented in the VASP code \cite{VASP}.  GGA+$U$ scheme was used to
describe the effect of correlation with the functional form proposed
by Liechtenstein et al. \cite{LDA+U1}; $U_{Ti}$=3 and 5 eV, and
$J_{Ti}$=0.5 eV. Lattice parameters of the bulk orthorhombic LTO are
shown in Table I. It is noted that the calculations with $U$=5
overestimate the experimental lattice parameters, while, with $U$=3,
the differences are reduced. We examined our system with the four
different sets: PBE or PBEsol and $U$=3 or 5 eV.  The changes caused
by these settings are found to be mostly quantitative.  The only
notable change was found in the rotation pattern of the
(LTO)$_2$/(LAO)$_2$ superlattice (see Sec.III.D).  In this case, the
relative stabilization energy among the various possible rotation
patterns ({\it i.e.} the relative rotation angles of oxygen octahedral
cages) is sometimes changed by $U$ value and the energy difference is
typically an order of a few meV per LTO/LAO.
%(The difference between stable conf with U=3 and stable conf with U=5
%is ~2meV per LTO/LAO when U=3 is used)
Importantly, for a given structure, the ground state spin and orbital
ordering patterns are found to be always the same. In this manuscript,
we mainly present the result of PBE and $U$=3 for the distorted
structure, as the Mott gap of LTO by optical spectra \cite{Okimoto},
$\Delta\sim$0.2eV, is in better agreement with the result of $U$=3
($\Delta\sim$0.56eV) than that of $U$=5 result ($\Delta\sim$1.88
eV). Since the gap is not opened with U=3 eV for the case of
tetragonal structure (with no distortion), we present U=5 result in
Sec.III.A.

We used the 2$a$$\times$2$b$ supercell, where the in-plane lattice
parameter is fixed to the STO value ($2a_0$=$2a$=$2b$=7.81\AA).  It
should be noted that the most stable distorted structure we found in
this study cannot be obtained with the $\sqrt{2}\times\sqrt{2}$
supercell.  To check the magnetic stability, we considered all possible
magnetic configurations and compared their total energies.\\[1mm]

\begin{table}
\begin{tabular}{l l c c c}
  \hline \hline
 System &  Type     &\ \ \   $a$ &\ \  $b$  &\ \ $c$  \\
  \hline
         & exp \cite{Cwik}  &\ \ \ 3.972 &\ \ 3.972 &\ \ 7.901  \\
         & PBE ($U$=3)      &\ \ \ 3.987 &\ \ 3.987 &\ \ 7.962  \\
Bulk LTO & PBE ($U$=5)      &\ \ \ 4.082 &\ \ 4.082 &\ \ 8.051  \\
         & PBEsol ($U$=3)   &\ \ \ 3.944 &\ \ 3.944 &\ \ 7.875  \\
         & PBEsol ($U$=5)   &\ \ \ 4.082 &\ \ 4.082 &\ \ 7.983  \\
  \hline
                    & PBE ($U$=3)      &\ \ \ 3.905 &\ \ 3.905 &\ \ 7.988  \\
(LTO)$_1$/(LAO)$_1$ & PBE ($U$=5)      &\ \ \ 3.905 &\ \ 3.905 &\ \ 8.025  \\
                    & PBEsol ($U$=3)   &\ \ \ 3.905 &\ \ 3.905 &\ \ 7.828  \\
                    & PBEsol ($U$=5)   &\ \ \ 3.905 &\ \ 3.905 &\ \ 7.862  \\
  \hline \hline
\end{tabular}
\caption{The optimized lattice parameters with the different $U$ values
  and the exchange correlation functionals.}
\end{table}

\section{Results and Discussion}

\subsection{Tetragonal structure}

\begin{figure}
\begin{center}
\includegraphics[width=6cm, angle=0]{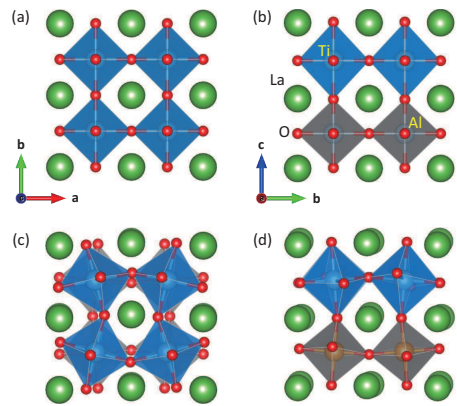}
\caption{ (Color online) The unitcell picture for (LTO)$_1$/(LAO)$_1$
  superlattice with tetragonal ((a) and (b)) and distorted structure
  ((c) and (d)). Top ((a) and (c)) and side views ((b) and (d)) are
  presented in which red and green balls represent the O and La atoms,
  respectively. The TiO$_6$ and AlO$_6$ cages are depicted in blue and
  brown, respectively.}
\label{structure}
\end{center}
\end{figure}

The bulk LTO has the lattice parameter of $a$=3.972 and the
GdFeO$_3$-type distortion ($Pbnm$), while LAO has $a$=3.778
\cite{Kumar} and $Imma$ under the tensile strain \cite{Hatt}.
Therefore, our system LTO/LAO can also have the structural distortion.
In this subsection, we first focus on the tetragonal structure without
such distortions to clarify their effects (see Fig.1(a)-(b)). In bulk
LTO, the lowest-$t_{2g}$ band is formed by the mixture of $d_{xy}$,
$d_{yz}$, and $d_{zx}$ orbitals, while the amount of their mixture
depends on the position of Ti atoms \cite{Solovyev-1,Khaliullin-2,
  Mochizuki,Pavarini,Solovyev-2}. On the other hand, in many Ti-based
superlattices, it is believed that the $d_{xy}$ band is in the lowest
energy due to the quantum confinement caused by heterostructuring.  A
recent X-ray absorption linear dichroism measurement for STO/LAO
\cite{JSLee} reported that the position of $d_{xy}$ band is lowered
and the $d_{yz}$/$d_{zx}$ degenerate in higher energy.  Previous
calculations (without considering GdFeO$_3$-like distortion) for
STO/LAO \cite{Pentcheva-STO/LAO} and LTO/LAO \cite{Seo} also concluded
that the lowering of the $d_{xy}$ band occurred due to the
confinement.

\begin{figure}
\begin{center}
\includegraphics[width=8cm, angle=0]{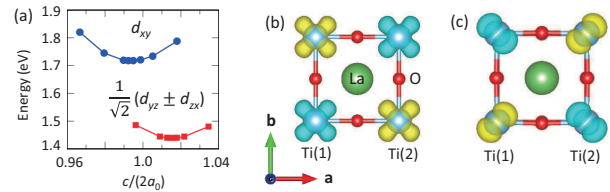}
\caption{ (Color online) (a) The calculated total energies (with $U$=5
  eV) of the tetragonal (LTO)$_1$/(LAO)$_1$ structure as a function of
  $c/(2a_{0})$, where $a_{0}$=3.905\AA \ is the in-plane lattice
  parameter of the STO substrate. The energy of the distorted
  structure (with $U$=5) is set to be 0eV.  (b)-(c) Spin density plots
  for the tetragonal (b) $d_{xy}$- and (c) $(d_{yz}\pm
  d_{zx})/\sqrt2$-phase. Yellow and blue represent different
  signs of magnetization.}
\label{structure}
\end{center}
\end{figure}

%Without structural distortion: c-sensitivity
However, as the spin, orbital, and structural degrees of freedom are
strongly coupled in these systems, the $t_{2g}$-band split in LTO/LAO
superlattice may not be as simple as previously reported.  Even in the
tetragonal phase, we found that the $t_{2g}$ band split is sensitive
to the in-plane and out-of-plane lattice parameters and the
lowest-lying $t_{2g}$ band can be reversed by strain.  In Fig. 2(a),
we present the calculated total energies as a function of $c/(2a_{0})$
and of the orbital configuration with $U$=5.  As the $c/(2a_{0})$
ratio varies, there are two minima in the total energy curve. At
around the first minimum of $c/(2a_{0})$=0.995
($c_{\textmd{LTO}}$=4.06 \AA \ and $c_{\textmd{LAO}}$=3.72 \AA), the
$d_{xy}^{1}$ orbital occupation is stabilized.  This is the state
obtained by the previous LDA+$U$ calculation \cite{Seo}, where the
lattice parameters have been obtained by the extrapolation from the
experimentally determined values of the unitcell and the geometrical
relaxation was performed within the cell. Fig. 2(b) shows the ground
state spin density of this $d_{xy}$-ordered phase.  AFM spin order is
stabilized over FM by 55 meV per LTO/LAO, which is consistent with the
conclusion of Ref.~\onlinecite{Seo}.

As the $c$ lattice parameter increases (within the tetragonal
symmetry; with no distortion), a new orbital ordered phase stabilizes
at $c/(2a_{0})$=1.015 ($c_{\textmd{LTO}}$=4.22 \AA \ and
$c_{\textmd{LAO}}$=3.73 \AA) in which the electron occupies
$(d_{yz}\pm d_{zx})/\sqrt2$ orbitals for Ti(1) and Ti(2), respectively
(see Fig.~2(c)). This antiferro OO is more stable than the previously
discussed ferro OO ($d_{xy}$) by 0.29 eV per LTO/LAO \cite{comm2}.  
The spin order is a checkerboard AFM, as shown in Fig. 2(c), which is more 
stable than FM by 6.8 meV per LTO/LAO.

The change of the orbital occupation as a function of $c/(2a_{0})$ can
be understood by examining the Ti--O bond lengths.  Since the STO
substrate imposes the in-plane compressive strain onto the LTO layer,
the out-of-plane Ti--O bond is elongated. As a result, the
out-of-plane bond length of the $(d_{yz}\pm d_{zx})/\sqrt2$-phase is
longer than that of the $d_{xy}$-phase by 0.08\AA, while the in-plane
bond lengths remain the same. The reduced Coulomb repulsion is, therefore,
responsible for the $d_{yz,zx}$ configuration becoming stabilized.\\[1mm]

\subsection{Distorted structure}

\begin{figure}
\begin{center}
\includegraphics[width=8cm, angle=0]{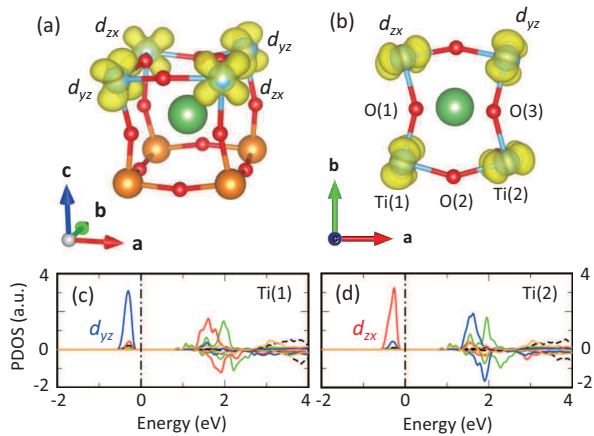}
\caption{ (Color online) (a)-(b) Spin density plots of the most stable
  distorted structure with antiferro OO.  (c)-(d) The projected DOS
  for (c) Ti(1) and (d) Ti(2).  Green, blue, and red lines represent 
  $d_{xy}$ $d_{yz}$, and $d_{zx}$ states, respectively.  Fermi energy
  is set to zero.}
\label{structure}
\end{center}
\end{figure}

As mentioned, considering the structural distortions in the bulk LTO
and the strained LAO, it is important to study the possible
distortions such as the oxygen octahedra rotations and their impact on
the electronic structure and magnetic property. From our total energy
calculations, the most stable structure of (LTO)$_1$/(LAO)$_1$ is
presented in Figs. 1(c)-(d) and Fig. 3(a)-(b), which is similar to the
$P2_{1}/m$ structure ($a^{+}b^{-}c^{-}$ in Glazer notation). This
structure is more stable than the tetragonal phase (with no rotation)
by 0.93eV per LTO/LAO with $U$=5 \cite{comm}.  The calculated band gap
by $U$=3 is 0.74 (Figs. 3(c)-(d)) while that of $U$=5 is 2.06 eV. The
$c/(2a_{0})$ of this phase is 1.023, which is longer than that of the
tetragonal $d_{xy}$-phase (see Fig. 2(a)).

Interestingly a different type of OO is found to be stabilized in the
distorted structure.  As presented in Figs. 3(a)-(b), $d_{yz}$ and
$d_{zx}$ orbitals are singly occupied at Ti(1) and Ti(2) site,
respectively.  Again, the ground state OO can be understood by
considering the bond length and the electrostatic energy: two in-plane
bond lengths, Ti(1)--O(1) and Ti(1)--O(2), are 2.063 and 1.970 \AA,
respectively, while the out-of-plane Ti--O is 2.135 \AA \ (see
Fig. 3(b)). Therefore, the $d_{yz}^{1}$ configuration at Ti(1)
minimizes the Coulomb repulsion between the electrons.  The same
argument holds for $d_{zx}^{1}$ at Ti(2) since the distances of
Ti(2)--O(2) and Ti(2)--O(3) are 2.073 and 1.964 \AA, respectively.

The ground state spin structure is also changed accordingly. Different
from the checkerboard AFM in the tetragonal phase, the FM spin order
is favored in this $P2_{1}/m$ structure having the less total energy
than AFM by 11.3meV per LTO/LAO.  The stabilization of FM spin order
is again well understood by the superexchange mechanism, which is
consistent with the Ti(1)-$d_{yz}$/Ti(2)-$d_{zx}$ OO. It is important
to note that, by making heterostructure, FM ground state has been
realized out of the AFM material, LTO. Our results demonstrate a
possible control of the ground state property. Figs. 3(c) and (d) show
the projected density of states (DOS) of Ti-$t_{2g}$ orbitals. One can
clearly see the bandwidths of $d_{yz}$ and $d_{zx}$ below Fermi level
are quite small $\sim$0.52 eV due to the two dimensional
confinement.\\[1mm]

\subsection{The effect of strain}

\begin{figure}
\begin{center}
\includegraphics[width=8.5cm, angle=0]{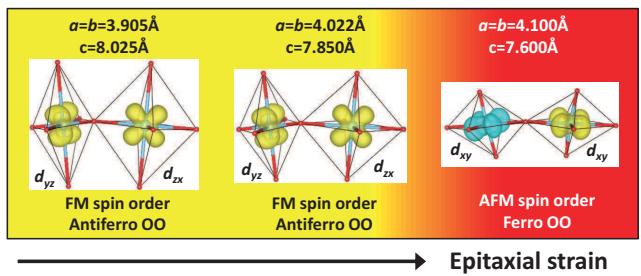}
\caption{ (Color online) Schematic phase diagram for the spin and
  orbital ground state as a function of epitaxial strain. Yellow
  represents the region of the FM-spin/antiferro-OO phase while
  red depicts the AFM-spin/ferro-OO. The phase transition occurs in
  between $a=4.022$ (+0.9\%) and 4.100 (+2.8\%)}
\label{structure}
\end{center}
\end{figure}

Considering the strong interplay between spin, orbital, and lattice
degrees of freedom, one may expect the possible control of the ground
state properties of this superlattice by epitaxial strain.  In order
to address this point, we performed the calculations with the three
different in-plane lattice parameters: i) $a=b=3.905$ \AA
\ (corresponding to STO substrate as discussed so far; $-$2.1\% of
compressive strain to LTO layer), ii) $a=b=4.022$ \AA \ (PrScO$_3$
(PSO) substrate; $+$0.9\% tensile strain), and iii) $a=b=4.100$\AA~
($+$2.8\% tensile strain) \cite{comm3}.

As schematically summarized in Fig. 4, the ground state spin and
orbital configuration is changed as a function of strain. While the FM
spin and antiferro OO is stabilized under the compressive and moderate
tensile strain, the AFM spin and ferro OO is stabilized
under a relatively large tensile strain of about +2.8\%.  
As the epitaxial strain increases
from $-$2.1 to +2.8 \%, the out-of-plane lattice parameter decreases.
The calculated Ti--O bond lengths along the out-of-plane direction are
2.135, 2.115, and 2.010 \AA \ for $-$2.1\%, $+$0.9\%, and $+$2.8\%
strains, respectively. At some critical strain, therefore, the
in-plane Ti--O bond length becomes longer than the out-of-plane one,
leading to a phase transition from the antiferro OO ( $d_{yz}$ and
$d_{zx}$) to the ferro OO ($d_{xy}$) as shown in Fig. 4.  The spin
order changes accordingly due to the superexchange interaction, and
the FM to AFM transition occurs simultaneously. For $a$=3.905 and
4.022 \AA, the calculated $\Delta E\equiv E$[AFM]$-$$E$[FM] (per
LTO/LAO) are 11.3 and 7.7 meV, respectively.  At $a=$4.100, $\Delta E$
becomes $-$7.5meV; that is, the checkerboard AFM is favored.

In the vicinity to the phase boundary, we found that a kind of mixture
of the two phases is stabilized. Our calculation with $a$=4.060 shows
that all $t_{2g}$ orbitals are occupied, but not equally for the four
Ti sites in the unitcell. The further details in nature of this phase
transition may require more study that is beyond the scope of
current work and our computation method.\\[1mm]

\subsection{The case of (LTO)$_2$/(LAO)$_2$}

\begin{figure}
\begin{center}
\includegraphics[width=6cm, angle=0]{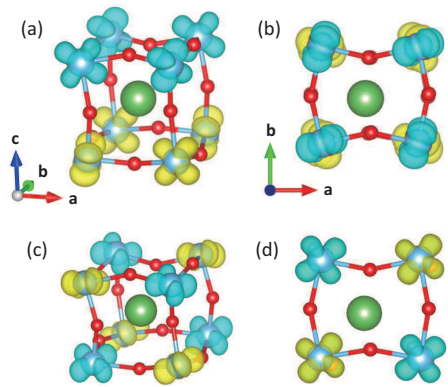}
\caption{ (Color online) The ground state spin density plots for the
  (LTO)$_2$(LAO)$_2$ superlattice calculated with (a)-(b) $a=3.905$,
  $a=4.100$. Note that the lower LTO layers are not seen by the upper
  LTO layers in (b) and (d).  Yellow and blue colors
  represent the up and down spin densities, respectively. }
\label{structure}
\end{center}
\end{figure}

In order to understand the thickness dependence we performed the
calculations for (LTO)$_2$/(LAO)$_2$. As the layer thickness
increases, more structural distortions can be stabilized. At
$a=3.905$, the most stable structure is the one in which two in-plane
rotation angles of TiO$_6$ octahedra are the same in their signs. The
ground state configuration of the spin, orbital, and lattice is shown
in Figs. 5(a)-(b). The spins are ferromagnetically aligned within the
$ab$-plane and antiferromagnetically along the out-of-plane
direction. As in the case of (LTO)$_1$/(LAO)$_1$, the
$d_{yz}$/$d_{zx}$ in-plane OO is accompanied with the structural
distortion and the spin order. Along the out-of-plane direction, the
$d_{yz}$/$d_{zx}$ occupation is alternating in such a way that the
occupied orbitals make some tilting angles between the layers, as
clearly seen in Fig. 5(b). This OO is found to be more stable
energetically than other types (with the same spin order) by 1.1--2.6
meV per LTO/LAO.

Under the tensile strain, the ground state configuration can be
changed. As in the case of (LTO)$_1$/(LAO)$_1$, at $a=4.100$ (the
tensile strain of +2.8\%), the $d_{xy}$ orbital is singly occupied and
the checkerboard AFM is stabilized within the $ab$-plane. Along the
out-of-plane direction, the spin order is AFM, and therefore G-type
spin order, is stabilized as shown in Fig. 5(c). The in-plane rotation
angles are also same in their signs, as shown in Fig. 5(d).\\[1mm]

\section{Summary}
The structural distortion and strain significantly change the ground
state spin and orbital configurations in LTO/LAO superlattice.  In
(LTO)$_1$/(LAO)$_1$, the FM and antiferro OO ($d_{yz}$ and $d_{zx}$)
phase is stabilized in a wide range of strain while the AFM and ferro
OO ($d_{xy}$) can be realized by applying a large tensile strain of
$\sim2.8\%$. Similar patterns of spin, orbital, and structural ground
state are also found in (LTO)$_2$/(LAO)$_2$. Our result not only
demonstrates the distinctive nature of the electronic and magnetic
properties in the TMO heterostructures but also shows a possible way
of controlling them by strain.\\[1mm]

\section{Acknowledgments}
We thank Ambrose Seo, Andy Millis, Harold Hwang and Heung-Sik Kim for
helpful discussions. This work was supported by the National Institute
of Supercomputing and Networking / Korea Institue of Science and
Technology Information with supercomputing resources including
technical support (KSC-2013-C2-005).


\begin{thebibliography}{99}
\bibitem{comm2} We note that this antiferro OO cannot be described
  with the 1$\times$1 unitcell, which is the part of the reason why
  previous studies found the $d_{xy}$ configuration as the ground
  state (Ref.~\onlinecite{Seo,Popovic,
    Okamoto-1,Pentcheva-LTO/STO}). With the same setting, we could
  also get this ferro OO.
\bibitem{comm} As mentioned in Sec.II, we mainly present $U$=3 eV
  result in this sub-section in order to present more realistic
  electronic structures to be compared with experiments.
\end{thebibliography}
\end{document}